# Tuning $La_{0.7}Sr_{0.3}MnO_3$ magnetic anisotropy via interfacial coupling in homometallic trilayers

N. Coppola[1], G. Vinai[2], A. Yu. Petrov[2], C. Sacco[1], H.S. Ur Rehman[1], S. Scarnicci[1], L. Capogna[3], C. Di Giorgio[2], C. Aruta[4], L. Maritato[1], A. Galdi[1]

[1]Dipartimento di Ingegneria Industriale, Università degli Studi di Salerno, Via Giovanni Paolo II 132, 84084 Fisciano (SA), Italy

[2]CNR – Istituto Officina dei Materiali (IOM), S.S.14, km 163.5, Trieste I-34149, Italy

[3]CNR-IOM Perugia, Dipartimento di Fisica e Geologia, Università di Perugia, Via Alessandro Pascoli, 06123 Perugia PG

[4]CNR-SPIN Via del Fosso del Cavaliere 100, 00133, Roma, Italy

## Abstract

In this work we explore the complex magnetic behaviour of $SrMnO_3/La_{0.7}Sr_{0.3}MnO_3/SrMnO_3$ trilayers combining magneto-optic Kerr effect (MOKE) and magneto-resistance measurements. The interplay of structural and orbital reconstruction and the competition of ferromagnetism/antiferromagnetism at the interfaces is responsible of a variety of complex magnetic properties, highlighting the use of heterostructures as an effective strategy to tune and control magnetic behaviour. We demonstrate the possibility to tune the magnetic properties of a ferromagnetic $La_{0.7}Sr_{0.3}MnO_3$ (LSMO) layer, embedded in two homometallic antiferromagnetic layers ($SrMnO_3$, SMO), by properly varying the role of interfaces using two different LSMO thickness values. Spin reorientation of the LSMO magnetic easy axis away from its room temperature orientation was observed below 125 K, due to the predominance of exchange interaction at the interface over the substrate-induced magnetic anisotropy, in agreement with established models for homometallic manganite heterostructures. The competition of super-exchange and exchange interactions at the interface results in a highly frustrated interfacial spin glass state, resulting in 20-fold enhancement of the coercive field and exchange bias. The modification of the magnetocrystalline anisotropy driven by the interfacial effects leads to a change in the anisotropic magneto-resistance response of the system, which dominates the low temperature behaviour.

## Introduction

Magnetic anisotropy, i.e. the directional dependence of the magnetic properties of a material, is crucial to the performance and functionality of magnetic materials used in devices such as magnetoresistive sensors with sub-nT sensitivity[1,2], magnetoresistive memories or logic devices.[3–5] Among perovskite oxides, the manganite family $La_{1-x}Sr_xMnO_3$ (LSMO) has attracted considerable attention owing to its rich interplay between spin, charge, and lattice degrees of freedom. Near optimal doping ($x \approx 0.3$), LSMO exhibits a ferromagnetic metallic state characterized by colossal magnetoresistance (CMR) near the Curie temperature and 100% spin-polarization at low temperature.[6] However, many phenomena besides CMR contribute to the magnetoresistance (MR) in LSMO, such as spin-dependent scattering at domain walls, grain boundaries, and anisotropic magnetoresistance (AMR) due to spin-orbit coupling.[7] AMR, which depends on the orientation of the magnetization with respect to the current direction, is most useful for the aforementioned applications, due to its low field response and switchable nature. Since it depends on the preferential alignment of the magnetic domains and on their rotation upon the application of magnetic field, AMR is naturally controlled by the magnetic anisotropy.

State-of-the-art thin film deposition techniques, such as pulsed-laser deposition and electron diffraction monitored shuttered layer-by-layer molecular beam epitaxy, have unlocked the capability of creating ordered, atomically flat heterointerfaces in functional oxide materials, where new interactions on the atomic orbital scale can occur.[8] In epitaxial heterostructures based on LSMO, the magnetic anisotropy can be engineered through strain, interface symmetry breaking, and layer thickness, providing a versatile platform for tailoring magnetic properties beyond bulk limitations.[9] Subtle modifications of epitaxial strain or interface bonding can induce transitions between in-plane and out-of-plane easy axes and can strongly influence the magnetoresistive response.[7,10]

$SrMnO_3/La_{0.7}Sr_{0.3}MnO_3/SrMnO_3$ trilayers (SMO/LSMO/SMO), in which a ferromagnetic (F) LSMO layer is embedded in two antiferromagnetic (AF) SMO layer, have been previously used to study the effect of electrostatic confinement of carriers in ultrathin LSMO layers.[11] In these homometallic structures (a term used herein to indicate that all layers share the same transition-metal cation, Mn, irrespective of their individual electrical transport properties), the LSMO layer is protected from the symmetry breaking at the air interface and from electrostatic or structural reconstructions at the substrate interface, often responsible for a "dead layer" of suppressed magnetic and transport properties,[12–14] becoming an ideal playground for exploiting interfacial coupling as a lever for tuning the AMR properties.

In this paper, we study the magnetic anisotropy (via magneto-optic Kerr effect, MOKE) and MR of SMO/LSMO/SMO trilayers grown on $NdGaO_3$ (NGO) (110) substrates, The LSMO thicknesses investigated are ≈ 3.5 nm and 5.8 nm (9 and 15 LSMO unit cells respectively).

While at room temperature the magnetic anisotropy of the LSMO closely resembles the one of LSMO/NGO (110) films, albeit with some dependence on the LSMO layer thickness, we observe significant changes at low temperature, when the SMO layer is AF. [14] These changes include the rotation of the F easy axis, exchange bias, and a reduction of the residual magnetization anisotropy, which consequently lead to a dominant role of AMR response at low temperatures in the trilayer structures. These results show how the activation of the AF properties of SMO at the interfaces of LSMO significantly affects the properties of the latter, making homometallic heterostructures valid candidates for AMR-based applications.

# Results

## Structural and electrical transport properties

SMO/LSMO/SMO trilayers were grown via shuttered layer-by-layer molecular beam epitaxy (MBE) on $GaO_2$-terminated NGO (110) substrates. The trilayers are nominally composed of a 15-unit cell (u.c.) thick SMO layer grown on top of the NGO substrate, a LSMO layer of variable thickness *n* u.c. and a 10 u.c. layer of SMO, as shown in the schematic Figure 1a.

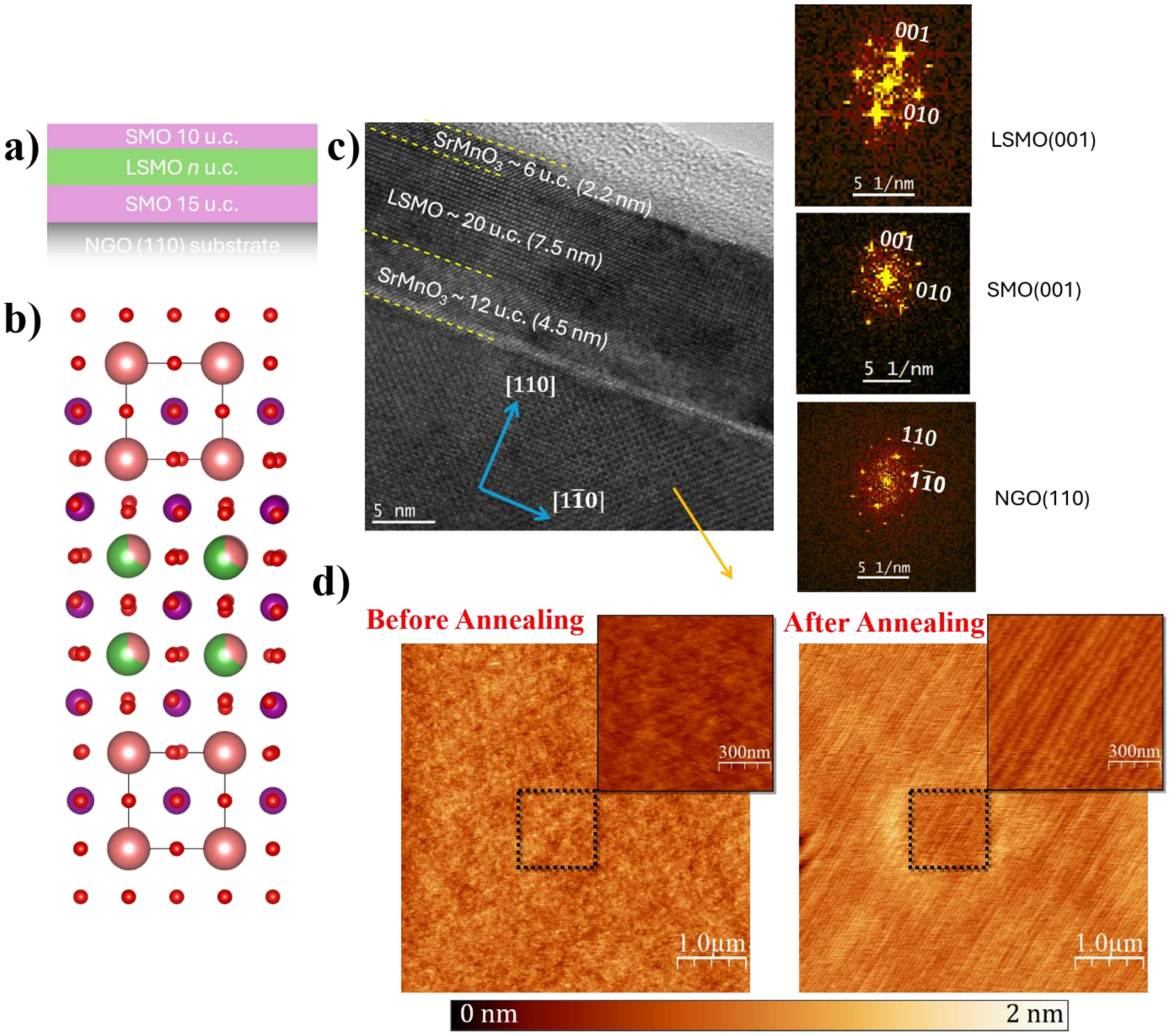

**Figure 1**: (a) schematic sketch of the trilayer stack; (b) schematic structure model of the SMO/LSMO/SMO interfaces (pink spheres: Sr, green: La, violet: Mn, red: O); (c) HR-TEM image of a sample with a 20 u.c. thickLSMO layer, where the insets display the corresponding digital Fast Fourier Transform (FFT) patterns calculated from the LSMO, SMO, and NGO regions, respectively; (d) atomic force microscopy of the NGO substrate before (left) and after (right) the annealing at 1000 °C in oxygen flow. The black dashed boxes indicate the area of the enlarged areas on the right.

An example of the NGO substrate surface, as-prepared and after the annealing treatment to achieve single termination, is shown in Figure 1d, which shows the improvement of the surface quality after the treatment (RMS roughness ≈ 0.8 nm), necessary for the high quality epitaxial growth of the trilayer.

All samples were routinely characterized via x-ray diffraction (XRD), which confirmed that the samples are fully strained on the substrate, and x-ray reflectivity (XRR) to check the thickness of the layers, as reported in ref.14. In bulk form, SMO possesses a high symmetry cubic perovskite structure with $Pm\bar{3}m$ space group ($a_{bulk}$=3.804 Å, $T_{Néel} \approx 230$ K), while LSMO, for $x = 0.3$ i.e. ferromagnetic with $T_C > 325$ K has a rhombohedral crystal structure (pseudocubic $a_{bulk} = 3.89$ Å) characterized by a pattern of octahedral tilts and rotations. The bulk crystal structures of SMO and LSMO [15–19] are depicted in the sketch of Figure 1b. When fully strained to the NGO(110) substrate (in-plane lattice parameters a = 3.864 Å, b = 3.854 Å), XRD measurements on 60 u.c. LSMO and SMO reference samples show that the out of plane lattice constant of LSMO expands to 3.91 Å and that of SMO contracts to 3.79 Å[14]. The discontinuity between the crystal symmetries and the substrate epitaxial strain is expected to induce a structural reconstruction at the SMO/LSMO interface; however, within the resolution of the HR-TEM measurements, we cannot detect significant deviation from a cubic structure in both the SMO and LSMO layers.[15–19]

however we could not detect any significant deviation from the cubic structure in both SMO and LSMO layers within the resolution of our HR-TEM measurements (Figure 1c).

Selected samples were measured via high resolution transmission electron microscopy (HR-TEM), which confirms the epitaxial growth and the realization of sharp atomically flat interfaces between the SMO and LSMO layers, , as shown in Figure 1c for the $n = 20$ trilayer. Conversely, the bottom NGO/SMO interface exhibits a slightly diffused character, as evidenced by a localized variation in the image contrast and a minor loss of atomic column sharpness across approximately 2–3 unit cells. This feature can be attributed to a minor chemical intermixing at the substrate boundary during the early stages of growth. The measurements show that the SMO layer acts as a defect scavenging barrier and effectively protects the LSMO layer from the interaction with interfacial and surface defects. The structural quality and epitaxial alignment are further supported by the Fast Fourier Transform (FFT) patterns shown as insets in Figure 1c. These digital diffractograms display well-defined diffraction spots corresponding to the pseudocubic LSMO (001) and SMO (001) phases, perfectly matching the (110) orientation of the NGO substrate. The measurement also confirms the nominal thickness of the LSMO layer (within 1 u.c.) and the average interplanar distance of ~3.9 Å. The interaction between different tilt and rotation patterns at perovskite interfaces is known to produce structural reconstructions that can significatively affect the material properties.

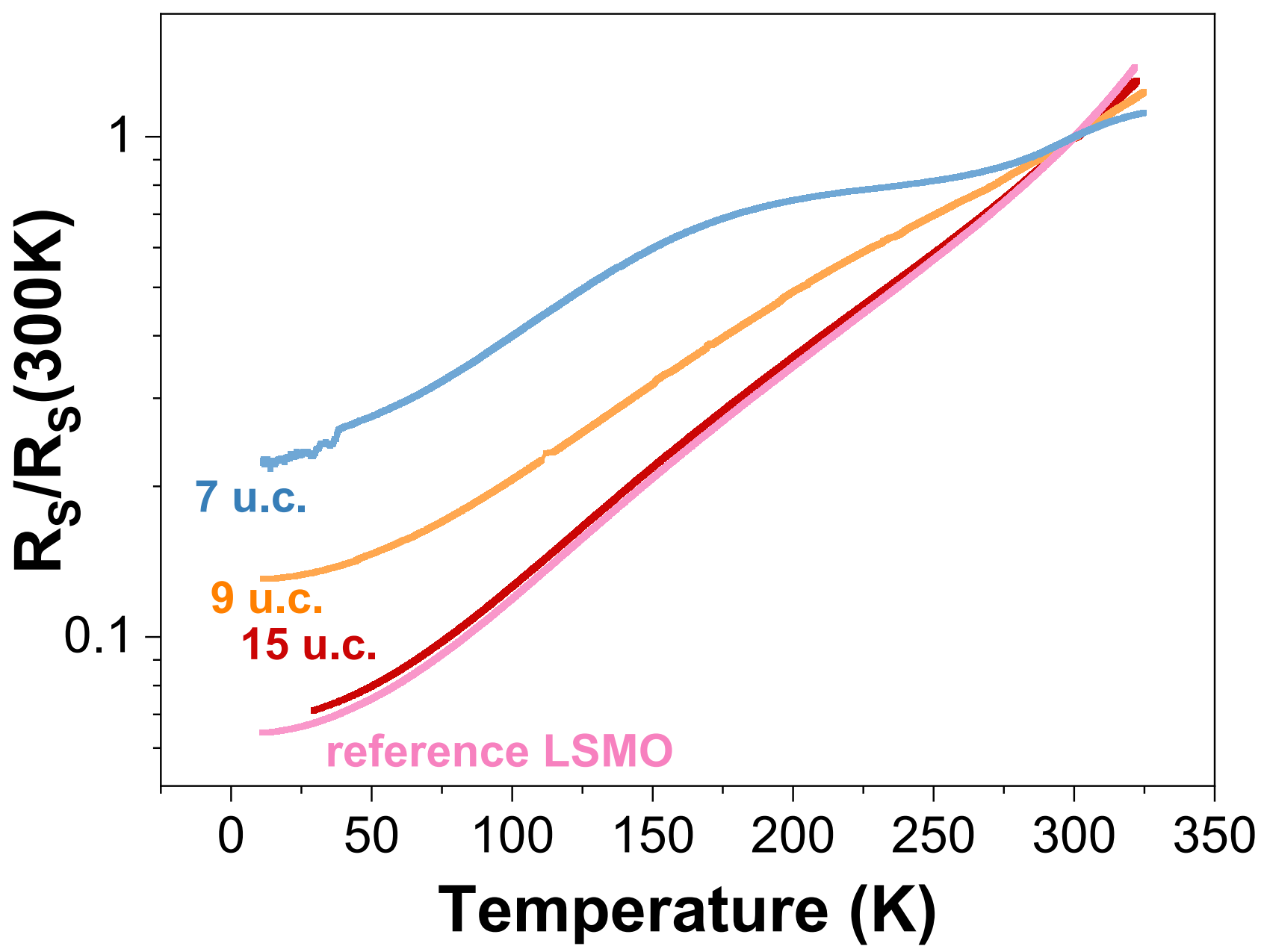

**Figure 2**: Sheet resistance ($R_s$) versus temperature of trilayers with different central LSMO thicknesses compared to a reference 60 u.c. thick LSMO film deposited with the same RHEED-assisted MBE growth protocol. The curves are normalized to room-temperature sheet resistance.

In Figure 2 we report the normalized sheet resistance $R_s$ as a function of temperature for the series of trilayers with varying number of LSMO u.c. *n* and of a LSMO single layer, which we use as reference, to show how the electronic transport properties are affected by the thickness. The metallic behaviour of the 60 u.c. reference LSMO thin film, with no metal-insulator transition up to 330 K (400 K if we consider the high temperature resistance measurements shown elsewhere for similar LSMO samples)[14] is in complete accordance with its phase diagram.[20] The trilayers with LSMO thickness $15 \leq n \leq 7$ are also characterized by metallic behaviour. We assume that the resistance of the trilayer is dominated by the LSMO layer, since the resistance of a pure (60 u.c.) SMO layer is too high to be measured in our setup. The residual resistivity ratio (RRR) at 10 K for $n = 15$ is close to the one of the LSMO reference sample and it decreases with decreasing *n* (*n=9 and 7*), while at $n = 4$ the behaviour is insulating. This change in electrical transport properties with thickness, as also observed elsewhere in trilayer heterostructures,[11,14,21] is attributed to the confinement of the carriers in the LSMO layer by the electrostatic potential well created by the SMO layers.[14]

## Magneto-optic Kerr effect measurements

MOKE measurements were made on the reference LSMO and on samples $n = 15$, *9* and *7* shown in Figure 2 (i.e. the metallic samples). However, the FM signal from the $n = 7$ trilayer, while detected, was too weak for proper analysis, so we focus our attention on samples $n = 15$ and *9*.

In Figure 3 we report the results obtained on the reference LSMO sample at both room temperature 300 K (i.e. above SMO Néel temperature $T_{Néel}$) and 77 K (well below it). In the following, directions in the sample plane are referred to the crystallographic directions of the NGO substrate.

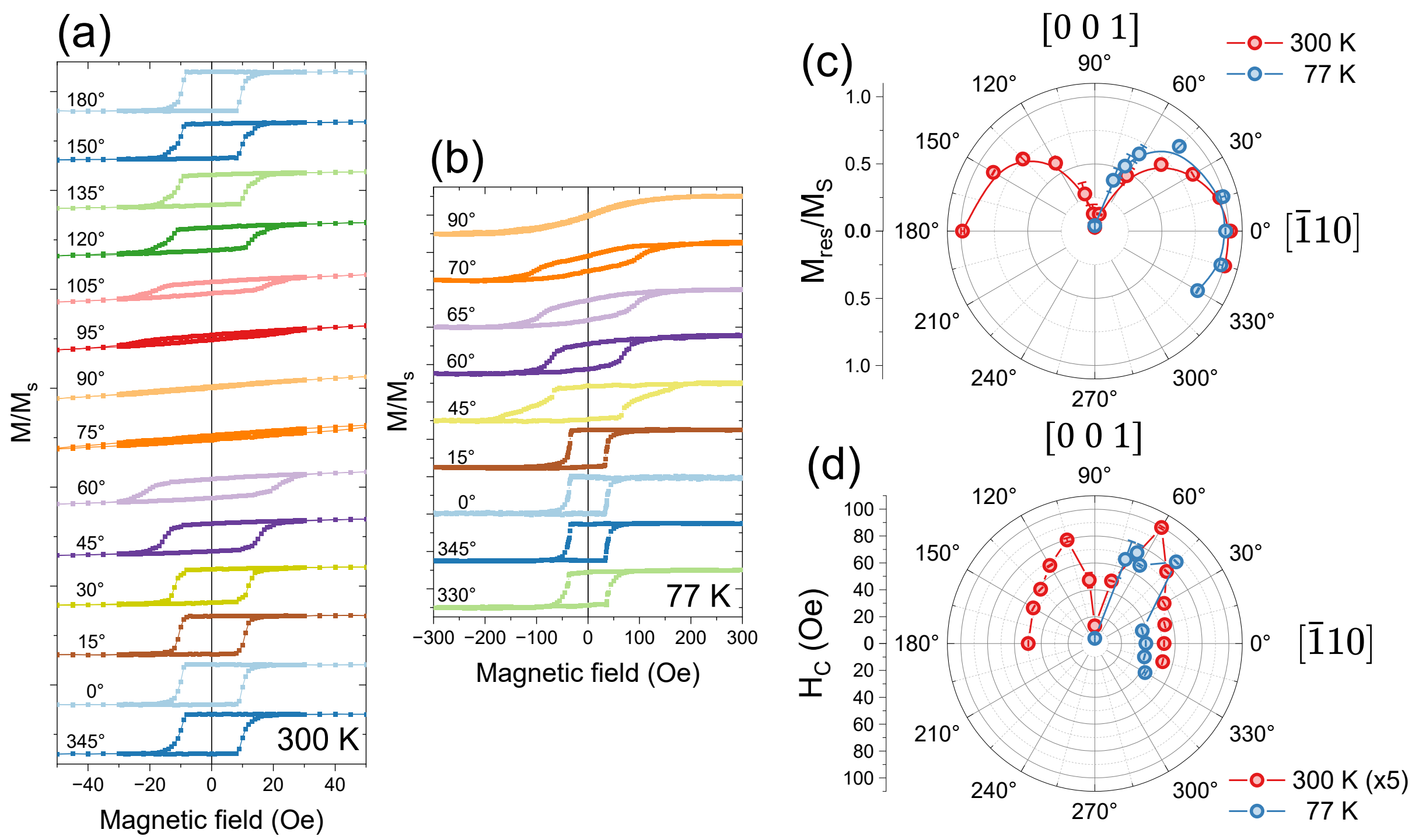

**Figure 3**: Rotational MOKE measurements on LSMO reference sample at 300 K and 77 K; a, b) hysteresis loops at 300 and 77 K respectively from which the polar plots shown in panels (c) and (d) are derived; c, d) polar plots of $M_{res}$ and of $H_C$ as a function of the orientation of the applied magnetic field **H** relative to the crystallographic axes of the NGO $[\bar{1}10]$ substrate.

The hysteresis loops reported in Figure 3a-b show uniaxial magnetic anisotropy character in the LSMO film consistently with what reported in [10] and [22-23] for LSMO films on NGO, where the uniaxial behaviour is attributed to the different length of the in-plane lattice constants imposed by the substrate. At both temperatures, the sample is characterized by a hard axis (HA) with zero residual magnetization ($M_{res}$) measured along NGO [001] in-plane axis, with the gradual increase of $M_{res}$ and coercive field ($H_C$) when approaching the easy magnetization axis (EA) along NGO $[\bar{1}10]$ in-plane axis. When cooled down at 77 K, $H_C$ is 4-5 times larger than the one observed at room temperature.

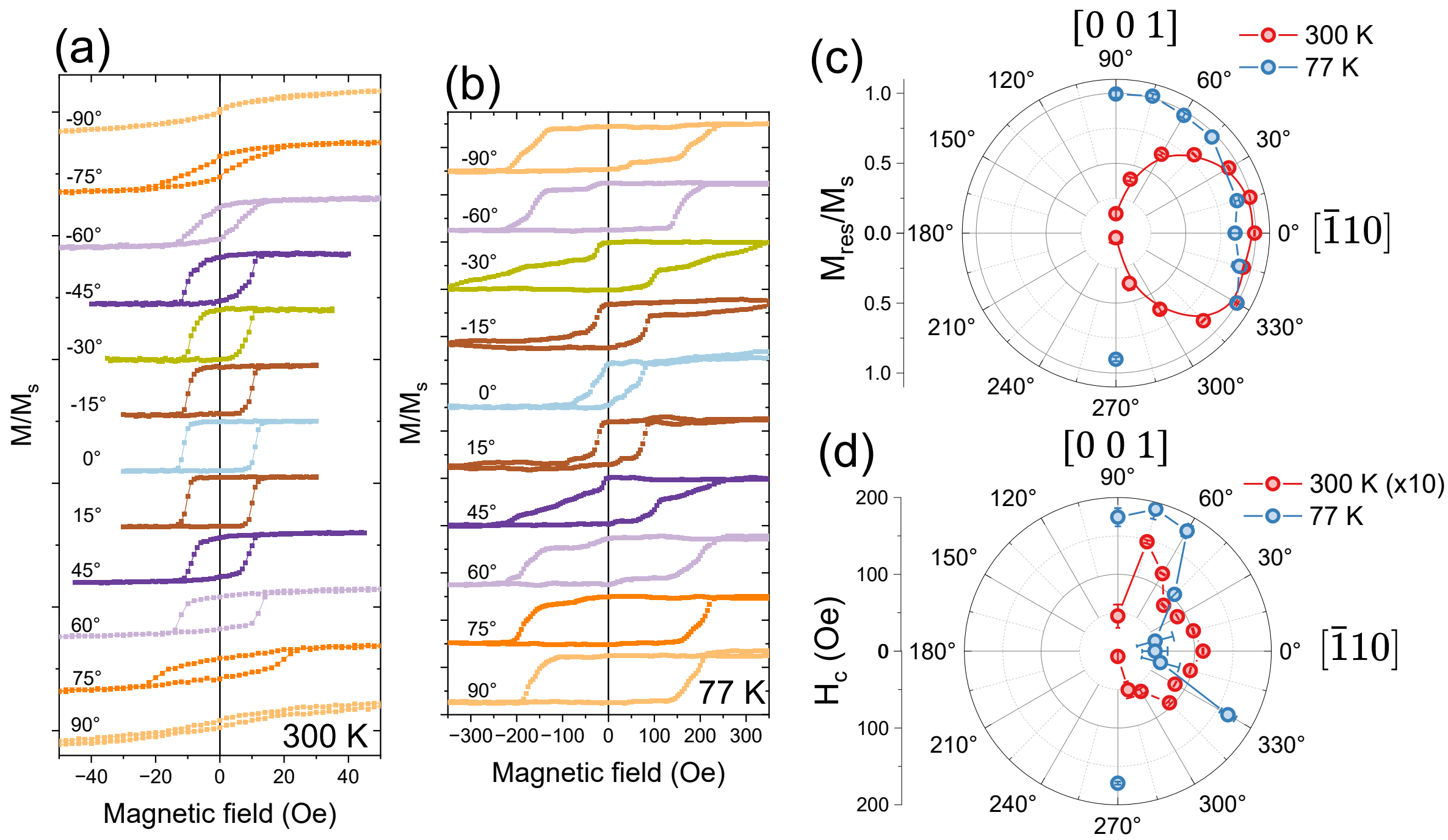

**Figure 4**: Rotational MOKE measurements performed on the $n = 15$ trilayer at 300 K and 77 K in ZFC condition; a, b) hysteresis loops at 300 and 77 K respectively from which the polar plots shown in panels (c,d) are derived; c) and d) polar plots of $M_{res}$ and of $H_C$ as a function of the orientation of the applied magnetic field **H** relative to the crystallographic axes of the NGO [110] substrate.

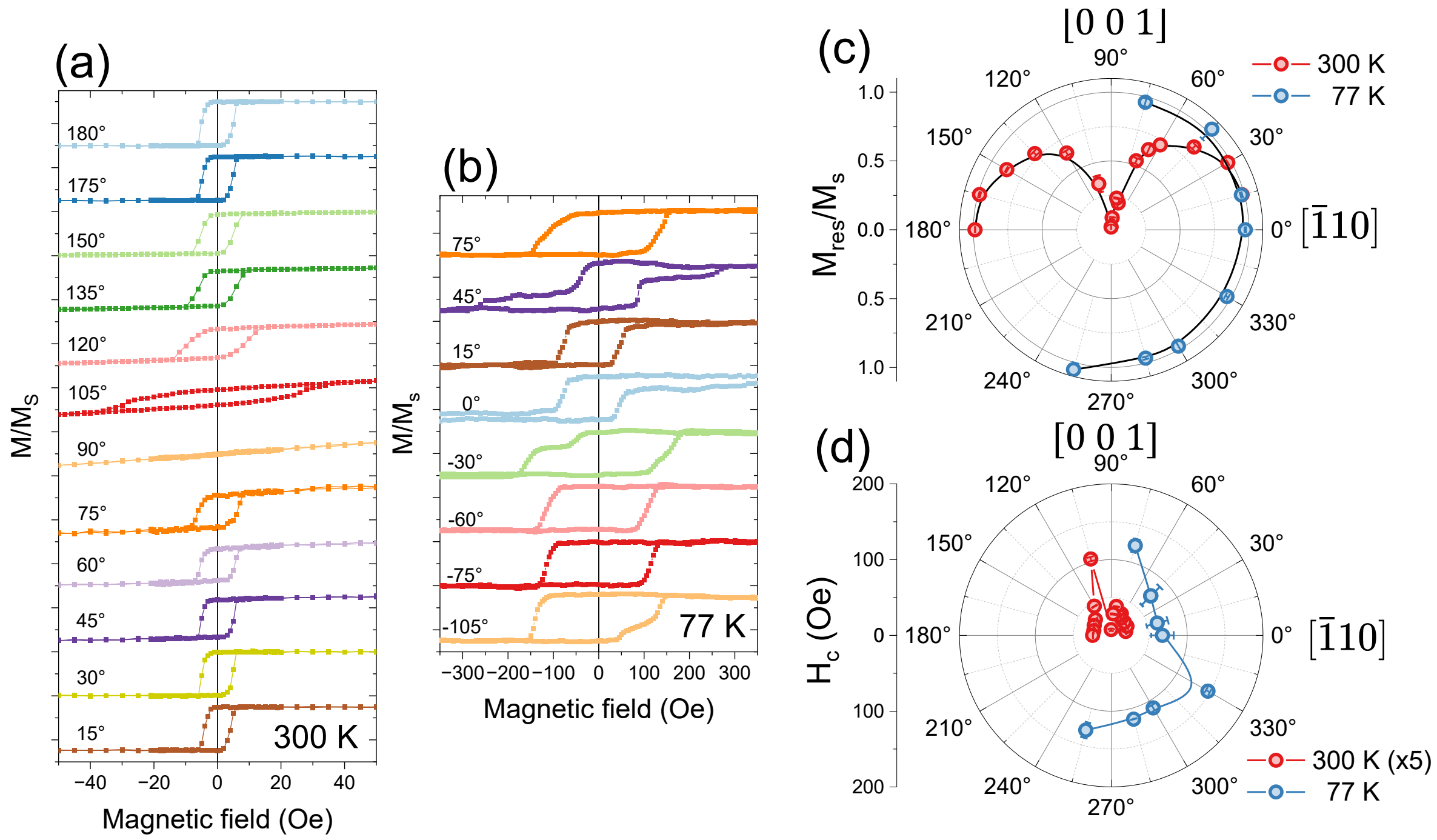

**Figure 5**: Rotational MOKE measurements performed on the $n = 9$ trilayer at 300 K and 77 K in ZFC condition; a, b) hysteresis loops at 300 and 77 K respectively from which the polar plots shown in panels (c,d) are derived; c) and d) polar plots of $M_{res}$ and of $H_C$ as a function of the orientation of the applied magnetic field **H** relative to the crystallographic axes of the NGO [110] substrate;

The room temperature hysteresis cycles of the *n=15* and *9* trilayers, reported in Figures 4a and 5a respectively, demonstrate that both maintain the uniaxial anisotropy of the LSMO reference sample, with EA along NGO $[\bar{1}10]$ direction. Magnetic hysteresis cycles at 300K for the trilayers along EA and HA are compared with those of the reference LSMO layer in Figure 6a and 6b. The resistivity and magnetization behaviour confirm that the 15 u.c. LSMO layer embedded in SMO behaves closely to the reference LSMO layer at room temperature, while the *n = 9* trilayer shows a reduction of both RRR and $H_C$. Finally, in Figure 6b we can observe that the anisotropy field $H_K$, i.e. the saturation field in the HA direction,[7] is smaller in the *n = 9* and*15* trilayers with respect to the reference LSMO.

The effects of the trilayer structure becomes more evident at temperature well below the $T_{Néel}$ of SMO (the AF order of the SMO layers in the trilayer heterostructures was verified in a previous work).[14] The hysteresis cycles in Figure 4b,5b measured after zero field cooling (ZFC), contain the contributions of different magnetic zones of the trilayers[14] and are the result of multiple interactions such as magnetoelastic, AF/F, magnetocrystalline and orbital reconstruction at the interface, so we cannot attribute the behaviour of $M_{res}$ (blue symbols in Figure 4c,5c) and $H_C$ (blue symbols in Figure 4d, 5d) to a single mechanism.

From Figures 4b,5b, we can observe that the hysteresis loops exhibit a more pronounced squared shape for angles that deviate from the $[\bar{1}10]$ direction. At 77 K the coercive field is significantly larger in both trilayers compared to the room temperature one, especially away from the $[\bar{1}10]$ direction which is the room temperature EA. As shown in Figure 6a,c,d, for the *n=15* trilayer, the $H_C$ increases by a factor of 4.4 with respect to room temperature along the $[\bar{1}10]$ direction, whereas it becomes 15 times larger along [001]. For *n = 9*, $H_C$ increases by a factor 14 at 77 K along the $[\bar{1}10]$ direction and by a factor 25 along [001]. Other characteristics of the 77 K hysteresis cycles are the presence of double step switching and asymmetry similar to exchange bias shift in ZFC measurements. The former is a signature of the complex magnetic anisotropy landscape in these samples, which may stem frommagnetization reversal through stable intermediate canted states (such as successive domain rotations). The latter will be discussed in the following. By comparing the angular dependence of $M_{res}$ and $H_C$ for *n = 15* (Figure 4c, d) at 77 K and 300 K, we can infer a switch of the EA by observing that the maximum of $M_{res}$ at 77 K is now along the [100] direction rather than along $[\bar{1}10]$, which also corresponds to the direction of maximum $H_C$. However, compared to 300K, the $M_{res}$ at 77K is only weakly modulated (by ~15%). For the *n = 9* sample, the $M_{res}$ (Figure 5c) has no detectable angular dependence and within our resolution is always equal to the saturation magnetization; however, the angular dependence of the hysteresis cycles and of $H_C$ are similar to those of *n = 15* (Figure 5b,d).

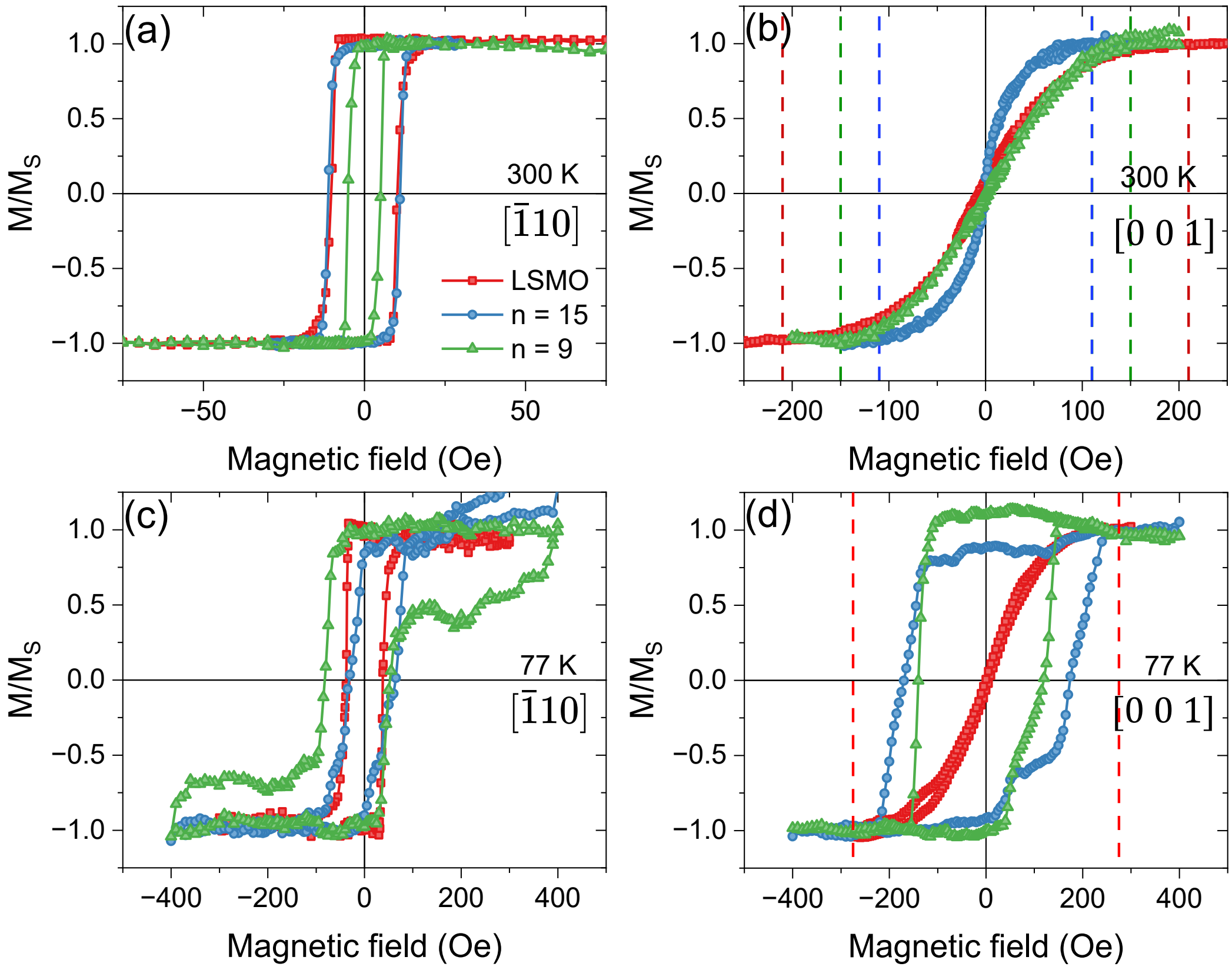

**Figure 6**: Comparison between the hysteresis loops of the three structures (red symbols: reference LSMO, blue: *n=15* and green: *n = 9* trilayers) along [$\bar{1}$10] and [001] NGO crystallographic directions at 300 K (a,b) and 77 K (c,d). All the loops were acquired via MOKE magnetometry in ZFC condition. Dashed lines in panels (b,d) indicate the anisotropy field $H_K$.[7]

The change of direction of the magnetic EA with temperature is a phenomenon known as spin reorientation transition (SRT), which has been observed in many LSMO-based heterostructures. [9,23–26] In order to find the SRT critical temperature, we measured the hysteresis cycle along the [001] direction at different temperatures, as shown in Figure 7a. We observe a change of the features of the hysteresis cycle from hard axis-like to easy axis-like at about 125 K. Antiferromagnetic order in SMO layer at low temperatures can induce exchange bias effect in LSMO film. Ding *et al.* in [27] pointed out that the arising of an exchange bias effect in LSMO/SMO bilayers is due to the onset (at ~180 K) of a spin-glass state due to the competition between AF double exchange and F superexchange interaction of Mn-O-Mn at the interface[27,28]. Below 130 K, the interaction between the LSMO F spins and the spin glass is strong enough to pin the LSMO spins in the FC direction, giving rise to exchange bias shift of the hysteresis cycles.

Here the setting of exchange bias was done on the *n = 15* trilayer by field cooling (FC) the sample down to 25 K under an in-plane magnetic field of +400 Oe applied along the [$\bar{1}$10] direction (Figure 7b) and along the [001] direction (Figure 7c).[27] Whereas the presence of double loops at certain angles is consistent with what observed in the ZFC hysteresis loops, we clearly observe an exchange coupling in Figure 7b, with the shift of the hysteresis loops with respect to the centre. Its negative sign is in accordance with what reported in literature on SMO/LSMO bilayers.[27,29] A much smaller effect is observed when the field is applied along the [001] direction, in Figure 7c. The comparison of ZFC and FC measurements reveals that the $[\bar{1}10]$ hysteresis cycle is shifted by about the same amount in both measurements. Since the LSMO is already ferromagnetic at 300 K, with **Mres** along $[\bar{1}10]$, the data indicate that even in absence of cooling field, this residual magnetization gets locked in place by the interaction with the interfacial spins, giving rise to the observed ZFC shifts at 77 K and 25 K (see Figure S1 of the Supplementary Materials).

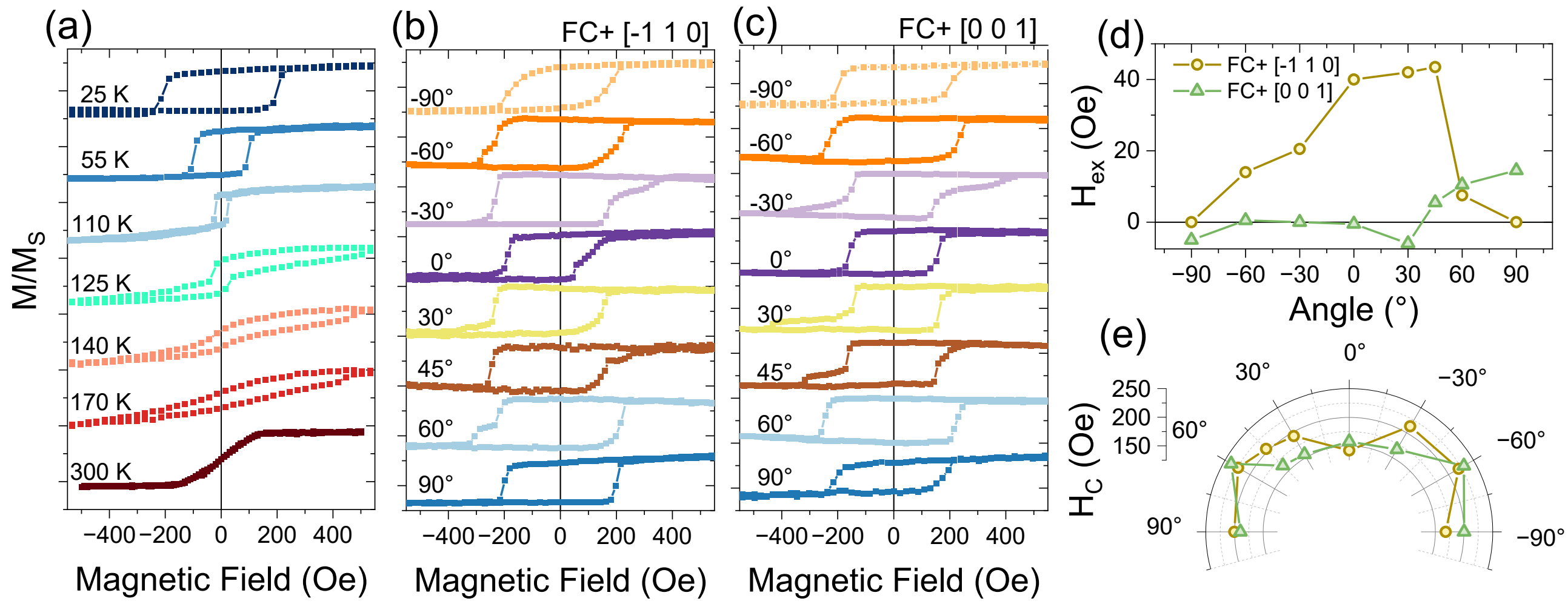

**Figure 7**: (a) Rotational MOKE measurements performed on the *n =15* trilayer acquired while lowering the temperature from 300 to 25 K. b,c) Rotational MOKE measurements performed on the n = 15 trilayer after field cooling with an in-plane magnetic field of +400 Oe (b) along the $[\bar{1}10]$ direction (0°) (c) along the [001] direction (90°); d,e) $H_{ex}$ and $H_C$ angular dependence. it.

## Magnetoresistance measurements

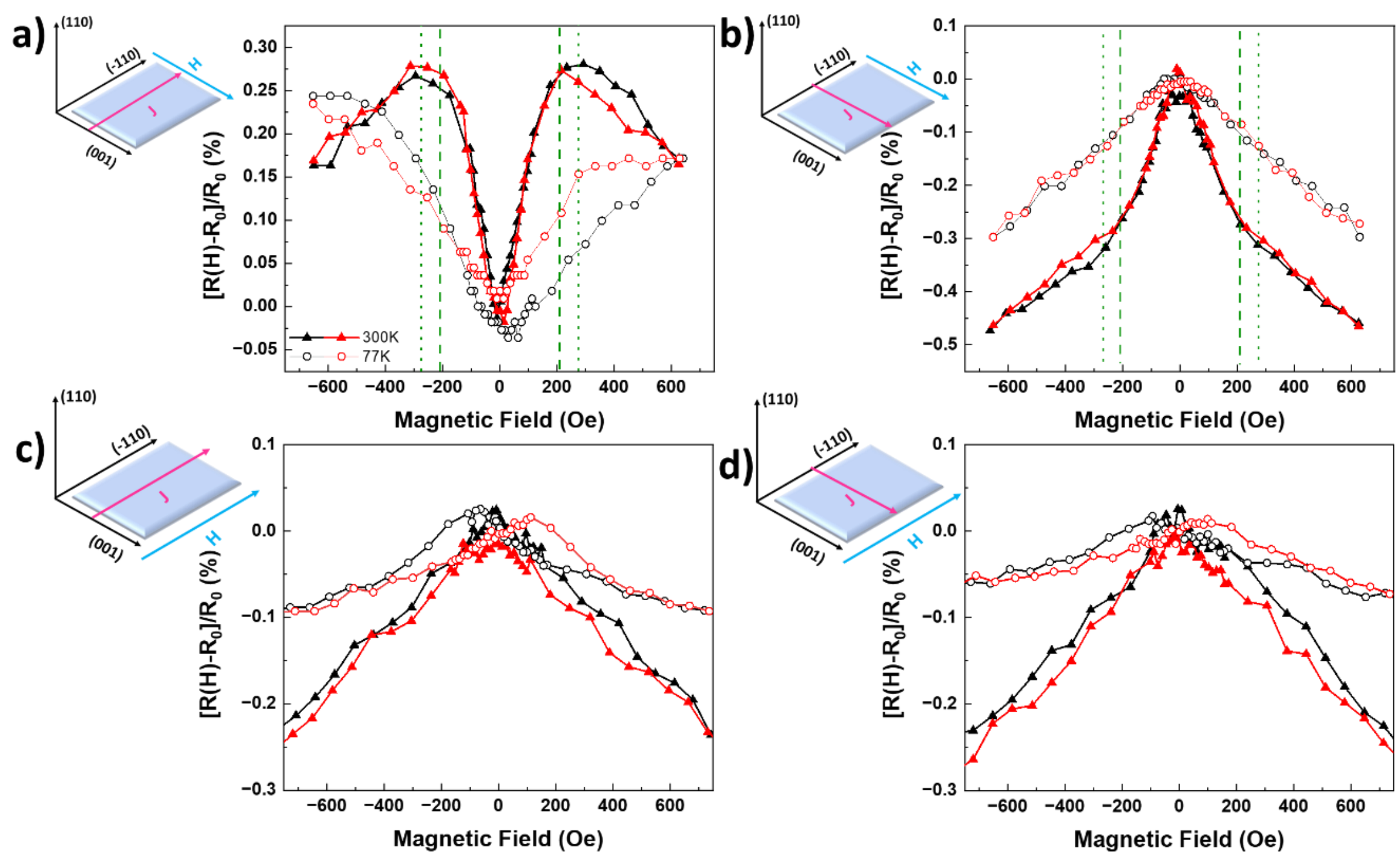

**Figure 8**: Magnetoresistance of the LSMO reference sample with the magnetic field (H) orthogonal (a,b) and parallel (c,d) to the EA, with the resistance measured with current (J) parallel (b,c) and orthogonal (a,d) to H. All values are plotted as % difference from the zero field resistance $R_0$. Black symbols represent the field sweep starting from positive values, whereas red from negative field values. Filled and open symbols represent 300 K and 77 K measurements respectively. Green dashed and green dotted lines mark the anisotropy field $H_k$ at 300 K and 77 K respectively.

One of the main consequences of the uniaxial magnetocrystalline anisotropy in LSMO thin films is the manifestation of a large AMR effect at low applied field, which has been observed on LSMO films grown on NGO and $SrTiO_3$, where uniaxial magnetic anisotropy is extrinsically induced by the vicinal substrates.[5,7,10] We therefore studied the MR of the *n = 15* trilayer sample at room temperature, where magnetically the LSMO layer behaves like the LSMO reference sample, and at 77 K, below the AF transition of SMO and the observed SRT of LSMO. For reference, we measured the MR of the LSMO reference, shown in Figure 8, demonstrating the typical signatures of the superposition of isotropic negative CMR and AMR. When the magnetic field (**H**) is applied along the EA $[\bar{1}10]$, the magnetization does not rotate in the sample plane, since the magnetic domains at zero field are already preferentially aligned along the $[\bar{1}10]$ direction in stripe domains of alternating magnetization, with a predominance of domains pointing in the $M_{res}$ direction.[7,10] Therefore, the current density (**J)** is always either perpendicular (Figure 8c) or parallel (Figure 8d) to both the magnetization direction and domain walls, so that the dominant effect is the negative CMR, with negligible AMR contribution. The MR curves in Figure 8(c-d), are characterized by a single negative slope which is larger at 300 K and smaller at 77 K, consistently with CMR. When the field is applied along [001] HA, the magnetization process involves the rotation of the magnetization from the $[\bar{1}10]$ to $[100]$ direction[7] and the tilting of the domain walls, so that the angle between **J**, the magnetization and the density of crossed domain walls changes.[10] This adds an AMR contribution to the MR, visible in Figure 8a,b, where two different slopes can be observed: a negative one at larger fields, due to CMR, and a second one at low field, negative when **J** is parallel to **H** and positive when orthogonal. The AMR contribution saturates at the anisotropy field $H_K$, *i.e.* the saturation field along HA that marks the completion of magnetization rotation (as reported in Figure 6b). From Figure 8a,b, we can observe that the AMR saturation value is about the same at 300 K and 77 K, while the wider range over which AMR is observed reflects the increase of $H_K$ Figure 6b,d).

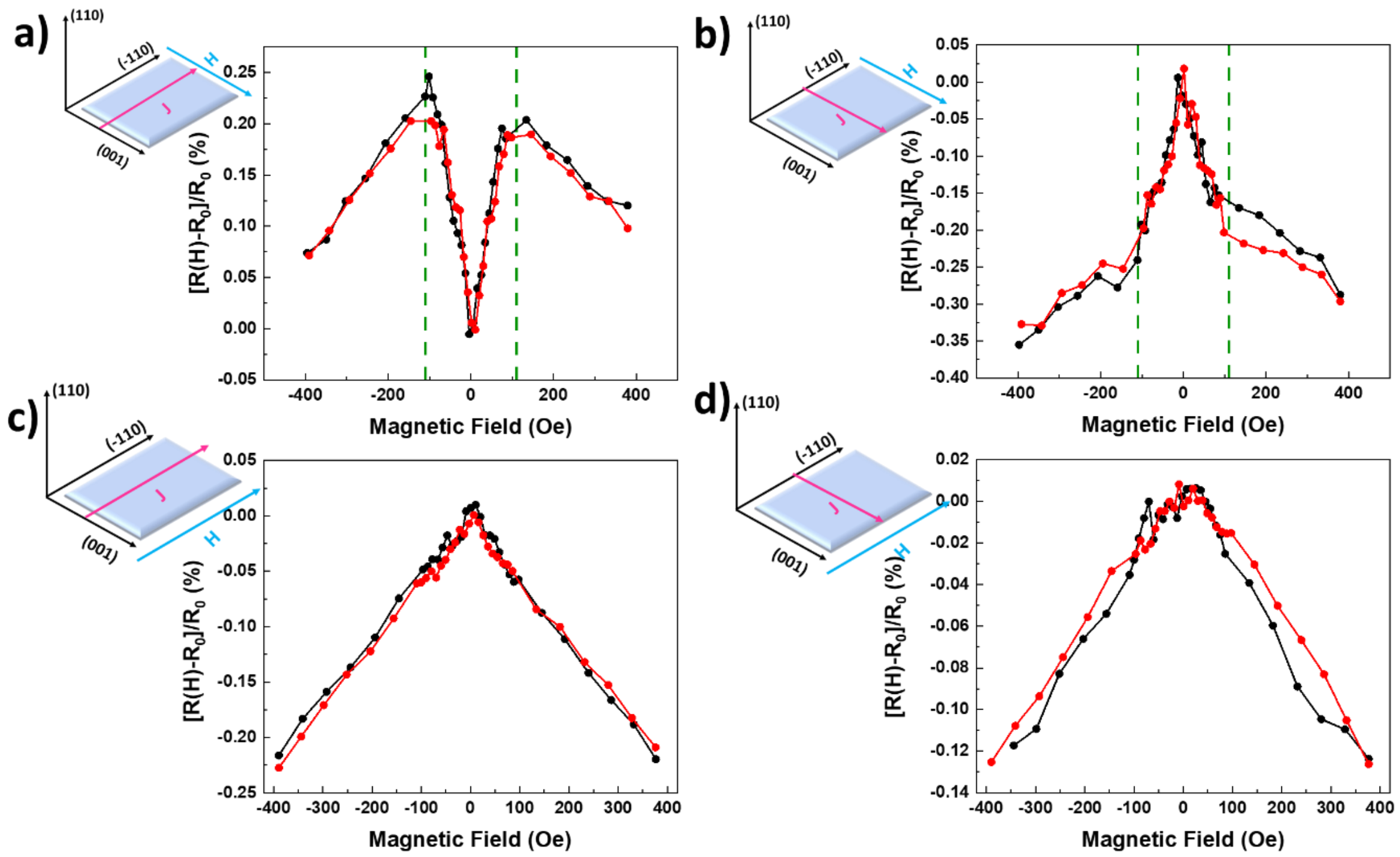

**Figure 9**: Magnetoresistance of *n = 15* trilayer measured at 300 K with the magnetic field (H) orthogonal (a,b) and parallel (c,d) to the EA, with the resistance measured with **J** parallel (b,c) and orthogonal (a,d) to **H**. All values are plotted as % difference from the zero field resistance $R_0$. Black symbols represent the field sweep starting from positive values, and red from negative ones. Green dashed lines represent the irreversibility field $H_k$.

The MR curves of the *n=15* trilayer, reported in Figure 9, reproduce the behaviour of the LSMO reference not only qualitatively, but also quantitatively in relative units. A significative difference is the width of the AMR region, which is reduced in the trilayer since it is characterized by a lower value of $H_K$, as evident in Figure 6(b). At 77 K (Figure 10), the behaviour of the MR of the trilayer is modified by the concurring SRT and AF/F interaction at the interface. The MR is dominated by AMR for both magnetic field orientations ([001] and $[\bar{1}10]$), since the MR is positive when $\boldsymbol{J} \perp \boldsymbol{H}$ (Figure 10(a) and (d)) and negative when $\boldsymbol{J} \parallel \boldsymbol{H}$ (Figure 10b,c). The splitting of the MR curves reflects the enhancement of the coercive field below $T_{Néel}$ in the trilayer samples shown in Figure 4(a), since the domains start rotating when H exceeds $H_C$, another signature of the AMR role. These observations are consistent with the nearly saturated value of $M_{res}$ at all angles reported in Figure 4(b), indicating that their rotation under applied field is the main mechanism responsible for the MR.

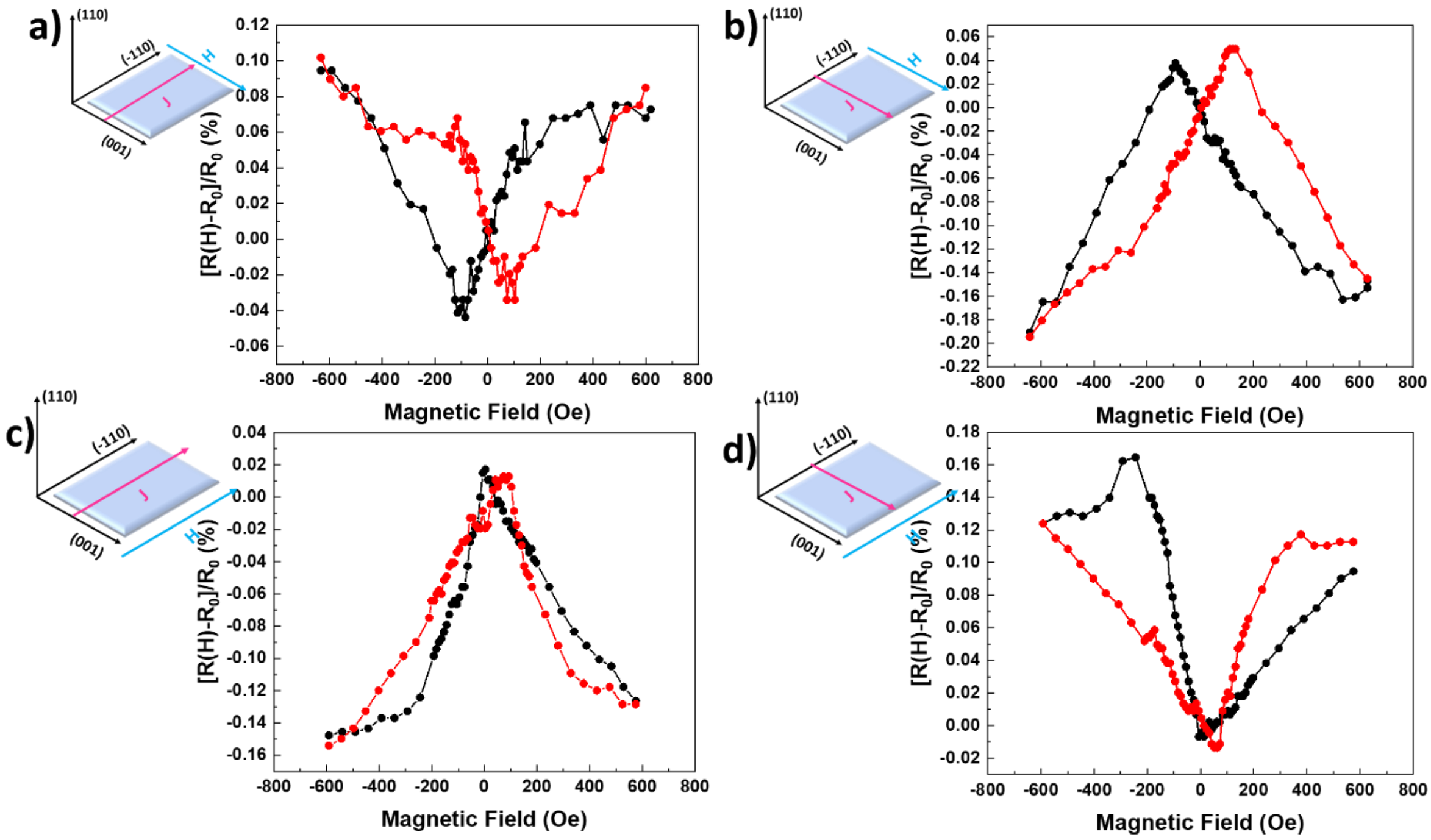

**Figure 10**: Magnetoresistance of $n$ = *15* trilayer measured at 77 K (ZFC) with **H** orthogonal (a,b) and parallel (c,d) to the EA direction $[\bar{1}10]$, with the resistance measured with **J** parallel (b,c) and orthogonal (a,d) to **H**. All values are plotted as percentual difference from the zero field resistance. Black symbols represent the field sweep starting from positive values, and red from negative ones.

# Discussion and conclusions

It has been shown that SMO/LSMO/SMO trilayers can preserve the metallic and FM properties of the LSMO layer at small thicknesses and reduce spurious effects occurring at the layer interfaces.[12,30,31] In our case, for instance, the bottom NGO/SMO interface exhibits a slightly diffused character over a few atomic layers in Figure 1(c) due to localized strain or intermixing, while the internal LSMO/SMO interfaces remain remarkably sharp and unaffected by this phenomenon. In LSMO epitaxial thin films on NGO (110), the propagation of the substrate tilts and rotations extends to the first 4 u.c. from the film-substrate interface [32], affecting the electronic structure of the first 6 u.c.[33]. Additionally, oxygen vacancy incorporation near the interface is also shown to affect the properties of the first few nm of film (7 nm in the case of LSMO on LSAT substrates) [34] .Our previous and current results show that a 15 u.c. thick homometallic (i.e. based on the same transition metal ion) buffer layer at the film-substrate interface [11,14] is capable of suppressing these effects. With the buffer and a top layer of sufficient thickness[11], ultrathin LSMO layers (15 and 9 u.c.) have transport and magnetic properties comparable at 300 K to those of a 60 u.c. LSMO layer. While at room temperature, the LSMO layer embedded in both $n$ = *15* and *9* heterostructures behaves similarly to LSMO single layer, showing the same magnetic EA, and the same coercive field at EA for $n$ = *15*. This suggests that at this temperature the baseline anisotropy is dominated by the epitaxial strain of the NGO substrate. Some signatures of softening of the ferromagnetic state with decreasing thickness can be evinced by the angular dependence of $H_C$ for *n = 15* and by the lower $H_C$ value of *n = 9*. However, substrate induced strain alone cannot justify the magnetic behavior at low temperature. At 77 K, we observe two concurrent phenomena in the trilayer samples: the rotation of the magnetic EA from $[\bar{1}10]$ to [001] (SRT), and an up to 20-fold increase of $H_C$ accompanied by a very weak angular dependence of $M_{res}$, which is nearly saturated in all directions.

SRT has been observed in several LSMO-based heterostructures, such as SMO/LSMO superlattices,[26] $LaCoO_3$/LSMO bilayers[25] and $La_{1-x}Sr_xCoO_3$/LSMO/ $La_{1-x}Sr_xCoO_3$ trilayers.[23] In such structures, SRT is deeply rooted in the localized electronic, structural and orbital reconstruction occurring at the interfaces. These phenomena are driven by polar discontinuities, charge transfer and symmetry mismatches (such as different octahedral tilts and rotations), coupled with the concurrent strain induced by the substrate and capping layers.[35,36] Specifically, it has been demonstrated that the energetic reordering of the Mn-3d orbital levels via interfacial charge reconstruction can directly trigger a macroscopic reorientation of the magnetic axes.[37] These interfacial effects become increasingly dominant upon lowering the temperature as the exchange interactions at the two interfaces become predominant with respect to the magnetic anisotropy induced by the substrate, causing a reorientation of the spins. Notably, this mechanism does not require the material at the interface with LSMO to be magnetic, and shown in [23,25,26], where no magnetic signal is detected from $La_{1-x}Sr_xCoO_3$ or SMO, probably because of their small thickness.

In our system, these features activate abruptly below 125 K rather than scaling monotonically with temperature as typically observed in systems displaying the SRT phenomenon.[27,38] This step-like emergence at a temperature comparable to the blocking temperature of SMO/LSMO interface[27,33] suggests that, while structural tilt reconstructions at the interfaces prepare the orbital landscape, the direct F/AF exchange interaction and magnetic coupling at the SMO/LSMO play an important role in determining the magnetization behavior.[37]

Indeed, in our heterostructures the SMO layers are AF at low temperature. Our group recently demonstrated via spectroscopic measurements that, while LSMO EA is aligned in plane along $[\bar{1}10]$ at RT and along [100] below the SRT, G-type AF SMO present its EA orientation out-of-plane.[14] As a consequence, at the SMO/LSMO interface the competition between F and AF is responsible of the occurrence of a new magnetic phase ordering characterized by a preferential out-of-plane orientation of the interfacial FM moments that extends to the SMO layer,[14] consistently with the emergence of a spin glass state at the interface.[28] The interaction between the FM spins of LSMO and the glassy phase at the LSMO/SMO interface is responsible for the increase of $H_C$ and for the setting of exchange bias coupling. The competition between the SRT and the F/AF interaction results in the behaviour of $M_{res}$, which is only weakly modulated as a function of angle in the *n = 15* trilayer, and always equal to saturation in $n = 9$, where the interface contribution prevails on the uniaxial anisotropy of the LSMO in the trilayer at 77 K.

The exchange bias measurements on $n=15$ also reveal that significant exchange bias is only observed when the setting field is applied along the room temperature EA direction (i.e. $[\bar{1}10]$), despite the occurrence of the SRT. Furthermore, the comparison between ZFC and FC measurements indicate that the presence of a non-zero $M_{res}$ (with the spins aligning preferentially parallel to the easy axis) above the blocking temperature is sufficient to give rise to the negative shift of the hysteresis cycle, without requiring saturation during cooling. These observations evince the competing effects of multiple interactions driving the magnetic anisotropy.

Finally, we verified the effect of this complex magnetic behaviour on the MR response of the $n = 15$ trilayer. At room temperature, the MR closely resembles the one of the LSMO layer; however, the trilayer structure enhances the AMR effect by affecting irreversibility field $H_K$. Consequently, the MR reaches 0.20% at 100 Oe, while in the LSMO reference layer it is 0.17% at the same field. At 77 K, the MR signal is dominated by AMR not only when the magnetic field is applied along the $[001]$ direction, but also when it is applied along the $[\bar{1}10]$ one. Since AMR is observed when the magnetization reversal process involves the rotation of the domains' magnetization,[7,10] this occurrence indicates that when the external field $\boldsymbol{H}$ is brought to zero, part of the domains prefers to align in a different direction than the one used to saturate the moments, for both $\boldsymbol{H} \parallel [\bar{\mathbf{1}}\mathbf{10}]$ and $\boldsymbol{H} \parallel [\mathbf{001}]$. The magnitude of AMR is independent of the field direction and only about 50-70% of the room temperature one, despite the much weaker modulation of $M_{res}$ with angle, thanks to the increased stability of the magnetization and mobility of the domain walls demonstrated by the squareness of the hysteresis cycles, as shown in Figures 4-5.[39,40]

In conclusion, SMO/LSMO/SMO trilayers display a complex magnetic behaviour due to the interplay of structural and orbital reconstruction and F/AF competition at the interfaces. Our experimental findings, which show a low-temperature SRT and a prominent exchange bias effect below RT, strongly support the physical scenario of interface-driven structural and orbital reconstructions coupled with the F/AF competition, as widely documented in literature for these manganite heterostructures. The magnetic properties can be tuned by varying the thickness of the LSMO layer, and we have also shown that this capability can be exploited to tailor the MR response of the system, in particular via exchange bias setting under field cooling and by maximizing the AMR contribution at low temperature instead of the CMR one. These results prove that homometallic trilayers are valuable candidates for AMR-based devices, paving the way for further investigations at different thicknesses regimes.

# Experimental Section

## Samples deposition

Thin film trilayers were deposited on NGO (110) single crystal substrates annealed at 1000°C in oxygen flow and then cooled down in the same atmosphere to ensure $GaO_2$ termination[41] and produce a terraced surface. The deposition is a RHEED (Reflection High Energy Electron Diffraction) monitored shuttered layer-by-layer MBE one with the substrate heated at 640°C in a 1.0 $10^{-6}$ Torr atmosphere. Each sample consists of a 15 u.c. ≈ 5.8 nm) $SrMnO_3$ (SMO) bottom layer and a 10 u.c. (≈ 3.8 nm) $SrMnO_3$ top layer in which a $La_{0.7}Sr_{0.3}MnO_3$ (LSMO) layer with varying thickness is embedded (see Figure 1). The central ferromagnetic LSMO layer was grown within a thickness series of $n = 4, 7, 9, 15,$ and *20* u.c. (corresponding to approximately 1.5, 2.7, 3.5, 5.8, and 7.7 nm, respectively).

$SrMnO_3$ and $LaMnO_3$ calibration samples were deposited monitoring RHEED oscillations to determine the right shutter opening times. Details of the deposition procedure can be found in [14]. Each trilayer deposition was followed by the deposition of a 60 u.c. LSMO reference to verify the stoichiometry and thickness accuracy during the growth.

## High Resolution Transmission Electron Microscopy (HR-TEM)

HR-TEM investigations were conducted using a JEOL JEM 2010 field emission (used source ZrO/W [100]) gun microscope operated at 200 kV T.A. The spherical aberration coefficient of the microscope Cs is 0.47±0.01 mm; the Scherzer resolution is 0.19 nm and the minimum probe size is 0.12 nm.

Before the measurement the sample was prepared making use of mechanical grinding and a precision ion polishing system (PIPS).

## Atomic Force Microscopy

Atomic Force Microscopy images have been acquired with a Veeco-AFM, equipped with Nanoscope V from Bruker, in the standard tapping mode technique, using a SCM-PIT tip from Bruker (Billerica, MA, USA). The cantilevers used were characterized by means of force curve spectroscopy and thermal methods, resulting in a resonance frequency and elastic constant of $f_0 \approx 75$ kHz and $k \approx 3$ N/m, respectively. The Atomic Force Microscopy images were treated to remove the artifacts due to the bending of the piezo scanner. The raw data were subjected to a standard background subtraction, and tilt removal, as well as to a de-spiking procedure, when needed. All the data have been analyzed by using the WSxM 5.0 software by NanoTec, 2015.

## MOKE measurements

Longitudinal Kerr Effect (MOKE) measurements were performed at the MBE Cluster of the NFFA-Trieste facility.[42] The signal was acquired using a s-polarized blue laser (405 nm) with a 500 µm diameter spot size for all samples; measurements were taken at 300 K in air and at 77 K with the sample in ultra-high vacuum. To ensure a high signal-to-noise ratio and verify the thermodynamic reproducibility of the magnetic states, all MOKE hysteresis loops presented in the main text (Figures 4, 5, and 6) were acquired by averaging over at least 10 consecutive measurement cycles. Polar plots were taken by rotating the samples along their azimuthal angle to measure the different anisotropic behavior as a function of orientation relative to the in-plane crystallographic axes [100] and $[\bar{1}10]$. The in-plane anisotropy was then evaluated by the angular dependence of $M_{rem}$ and $H_C$. The coercive field was evaluated after averaging the half width hysteresis loops at zero magnetization for both negative and positive sides of the loop.

### Resistance vs temperature and magnetoresistance measurements

The resistance and magnetoresistance measurements were performed with a Janis CCS-300S closed-cycle cryostat coupled to a compressor (Sumitomo HC-4E1 Helium Compressor, Sumitomo (SHI) Cryogenics of America, Inc., USA) at temperatures between 10 and 325 K, using 4 contacts Van Der Pauw configuration. For magnetoresistance measurements, the cryostat was inserted in the expansion of an electromagnet with Helmholtz coils configuration (Lakeshore EM4-HVA). The magnet current was controlled via a programmable power supply, and the value of the generated magnetic field was measured with a Hall probe placed next to the cryostat head (Lakeshore 425 Gauss meter and HGT-3010 sensor). The magnetic field was applied in the sample plane either parallel or perpendicular to the EA, while the current direction was changed via the contact configuration.

## Acknowledgements

This work has been performed in the framework of the Nanoscience Foundry and Fine Analysis (NFFA-MIUR Italy Progetti Internazionali) facility. The authors wish to acknowledge the contribution of Dr. Piu Rajak for performing the HR-TEM measurements.

# Supplementary Materials

# Tuning $La_{0.7}Sr_{0.3}MnO_3$ magnetic anisotropy via interfacial coupling in homometallic trilayers

N. Coppola[1], G. Vinai[2], A. Yu. Petrov[2], C. Sacco[1], H.S. Ur Rehman[1], S. Scarnicci[1], L. Capogna[3], C. Di Giorgio[2], C. Aruta[4], L. Maritato[1], A. Galdi[1]

[1]Dipartimento di Ingegneria Industriale, Università degli Studi di Salerno, Via Giovanni Paolo II 132, 84084 Fisciano (SA), Italy

[2]CNR – Istituto Officina dei Materiali (IOM), S.S.14, km 163.5, Trieste I-34149, Italy

[3]CNR-IOM Perugia, Dipartimento di Fisica e Geologia, Università di Perugia, Via Alessandro Pascoli, 06123 Perugia PG

[4]CNR-SPIN Via del Fosso del Cavaliere 100, 00133, Roma, Italy

## S1. Positive and negative field cooling loops

In the following figure (Figure S1) hysteresis loops measured in ZFC, +FC (+400 Oe) and – FC (-400 Oe) conditions are shown.

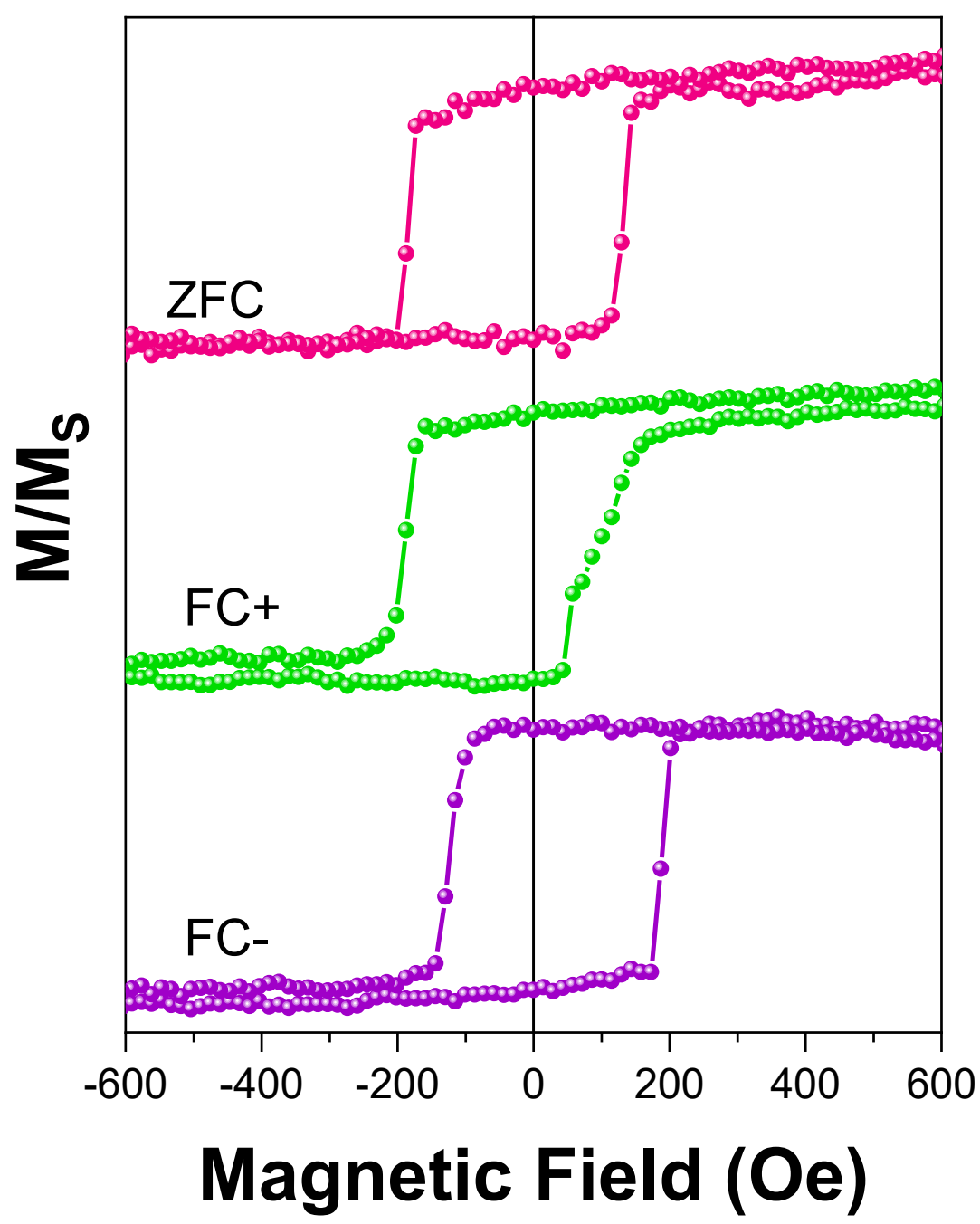

**Figure S1**: hysteresis loops measured for the sample $n = 15$ along the $[\bar{1}10]$ direction in ZFC condition (pink line), in +FC condition (+400 Oe, green line) and in -FC condition (-400 Oe, violet line).

All the three loops were acquired in identical alignment conditions and confirmed a symmetric reversal of the horizontal loop shift as can be seen in the figure above, which reports the hysteresis curves along the $[\bar{1}10]$ direction. The figure also shows that the ZFC curve is characterized by almost the same shift as the FC+ curve. The reason is that the LSMO layer is characterized by a $T_C > 300$ K and possesses a non-zero residual magnetization aligned along the easy axis at room temperature (i.e. the $[\bar{1}10]$ direction). This magnetization remains set once crossing the blocking temperature of the system even in absence of setting field. In case of ZFC, the sign of exchange bias depends on the previous magnetic history of the sample. In our typical measurement protocol, the measurement of the hysteresis cycles always ends with positive field, thus leaving the sample with positive magnetization.